\documentclass[twocolumn,showpacs,preprintnumbers,amsmath,amssymb]{revtex4}
\usepackage{amssymb}

\usepackage{graphicx}
\usepackage{dcolumn}
\usepackage{bm}
\usepackage{color}

\begin{document}



\title{Spectroscopy of Superradiant scattering from an array of Bose-Einstein Condensates}

\author{Xu Xu}
\author{Xiaoji Zhou}\thanks{Electronic address: xjzhou@pku.edu.cn }
\author{Xuzong Chen}

\address{School of Electronics Engineering $\&$ Computer
Science, Peking University, Beijing  100871, China}

\date{\today}


\begin{abstract}
We theoretically study the superradiant gain and the direction of
this coherent radiant for an array of Bose-Einstein condensates in
an optical lattice. We find that the density grating is formed to
amplify the scattering light within the phase match condition. The
scattering spectroscopy in the momentum space can provide a method
for measuring the overlap of wavefunction between the neighboring
sites, which is related to their inner-site and inter-site
coherence.
\end{abstract}


\maketitle

\section{Introduction}
Superradiance from a Bose-Einstein Condensation (BEC) offers the
possibility to study the novel physics associated with cooperative
scattering of light in ultracold atomic systems. A series of
experiments~\cite{Inouye1999science, Schneble2003scince, 1999, 2004,
Schneble2006prarc, KMRvdStam2007arxiv, Bar-Gill2007arxiv, li} and
related theories~\cite{Moore1999prl, Zobay2006pra, Pu2003prl,
Mustecaplioglu, Robb2005job, Uys2006arxiv} have sparked the related
interest in the quantum information ~\cite{KMRvdStam2007arxiv},
collective instability~\cite{Courteille1, Courteille2}, high
precision measurement~\cite{wineland}, and coherent atom
optics~\cite{Schneble2003scince, sadler}.

In a typical BEC superradiant experiment,the pattern of recoiling
atoms by absorption image method reflects the atomic momentum
spectroscopy or momentum distribution~\cite{Schneble2003scince,
1999, 2004, Schneble2006prarc, KMRvdStam2007arxiv,
Bar-Gill2007arxiv, li}, where the moving atoms and the static BEC
form a matter wave grating. At the same time, the scattering optical
spectroscopy shows the gain process with
time~\cite{Inouye1999science}. To enhance the scattering light
signal, the optical cavity is applied in the similar experimental
setting which usually called collective atomic recoil lasing
(CARL)~\cite{Courteille1, Courteille2,Bonifacio}, where the atoms
are forced to maintain in the density grating by the optical lattice
in the cavity. Different from the above case, here we consider the
scattering superradiance from an array of BECs.

Optical lattices (OL), created by pairs of off-resonance
counter-propagating laser beams, offer new opportunities to
investigate quantum information processing and strongly correlated
quantum matter~\cite{bloch}. The periodical potential in an optical
lattice forms an atomic density grating, hence to study the
superradiance in this array the coherence of atom both inner site
and between sites need to be considered. Therefore superradiance has
the potential to become a method to detect the coherence of atoms in
an optical potential. This is different from the interaction of
light and BEC in an OL trap without the atoms'
recoiling~\cite{Mekhov}, where the inter sites atomic coherence was
considered and the inner site coherence is neglected.

To study the superradiance in an optical lattice, there are several
problems to be considered with regard to the theory about the
superradiance from BEC~\cite{Moore1999prl}. First the frequency of
optical lattice, usually several $kHz$, is much larger than that of
the magnetic trap (tens to hundreds $Hz$) where the effects of trap
is usually neglected for its frequency is much smaller than the
recoil frequency. Secondly, we need to calculate the gain in the
special emission angle. In the case of a magnetic trap, the atomic
cloud experiences the maximum gain when the mode is along its long
axis. And it can also be understood by that the direction is
selected for the least width of momentum to get the maximum
gain~\cite{Moore1999prl}. The long axis and the least width of
momentum, these two directions are the same in a magnetic trap, but
not necessarily true in an OL trap. Lastly but important,since the
light scattering depends on the coherence of different sites, the
interference of scattered light results in amplification at some
specific frequency and suppression at the others. It could provide
us a method to obtain the information of the atoms in OL. The
density grating formed by optical lattice and the grating formed by
moving and static atoms give two criterions for the optical
amplification. Hence, similar to the coherent-enhanced imaging where
Raman superradiance is used to probe the spatial coherence of BEC in
a magnetic trap~\cite{sadler}, the scattering spectroscopy reflects
the cooperative radiation of atoms inner-site and inter-sites.

\section{Gain for the Condensate in the Trap}

\begin{figure}
  \includegraphics[width=6cm]{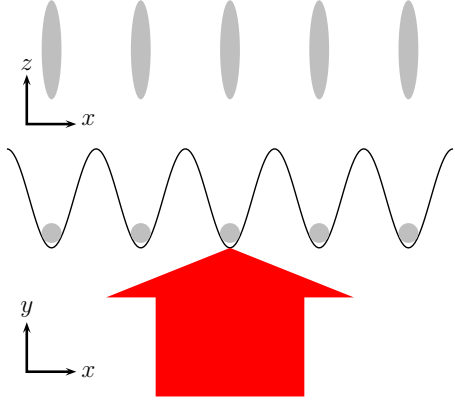}
  \caption{(Color online) The system sketch. For a cigar shaped condensate,
  the optical lattice is formed along $x$-axis, with M sites. The
  pumping laser is propagating along $y$-axis. And in the $x-z$ plain,
  the length in the $z$-direction is bigger than that in the $x$-direction. }
  \label{fig:1}
\end{figure}

We consider the model that the atomic cloud is prepared in an
optical lattice, as shown in Fig.~\ref{fig:1}. The optical lattice
is placed along the $x$-axis with $M$-sites centered at origin. The
lattice constant is $a_0=\lambda/2$, the length of condensate is
$L=Ma_0$, and the pumping laser incident with wave vector
$\mathbf{k}_{0}$ propagates along its short axis $y$.

In order to investigate the superradiant gain, we adiabatically
eliminate the excited state for the far-off resonant pump laser and
use the rotating wave approximation. Therefore the effective
Hamiltonian about the coupling between the atomic and
electromagnetic fields can be written as~\cite{Moore1999prl}
\begin{equation}
 \hat H = \hat{H}_a + \hat{H}_p+ \hat H_{i},
\end{equation}
where the $\hat H_a = \int
\rm{d}^3\mathbf{r}\hat\Psi^\dagger(\mathbf{r})[\hat p^2/2m+ \hat
V(\mathbf{r})]\hat\Psi(\mathbf{r})$ is the atomic Hamiltonian, for
the classical potential of OL is $V(\mathbf{r}) = V_0\cos(x/a_0)$
with the lattice depth $V_0$. $\hat H_p = \int
\rm{d}^3\mathbf{k}\hbar\omega_k \hat b^\dagger(\mathbf{k})\hat
b(\mathbf{k})$ is the Hamiltonian of photon, and the interaction
Hamiltonian is
\begin{equation}
   \hat H_i=\int d^{3}\mathbf{k} d^{3}\mathbf{r}[\hbar g(\mathbf{k})\hat{\Psi}^\dagger(\mathbf{r}) \hat{\mathbf{b}}^\dagger
                (\mathbf{k}) e^{i(\mathbf{k}_{0}-\mathbf{k}) \cdot \mathbf{r}} \hat{\Psi}(\mathbf{r}) + H. c.],
\end{equation}
where $\hat{\Psi}(\mathbf{r})$ is atomic field operator, and
$\hat{\mathbf{b}}^\dagger(\mathbf{k})$ is the annihilation operator
for a photon in mode $\mathbf{k}$ in the frame rotating at the pump
frequency $\omega_{0}$. Here the photon energy $\omega_\mathbf{k}=c
|\mathbf{k}|-\omega_{0}$, and $g(\mathbf{k})$ is the coupling
coefficient for scattering between the pump and vacuum modes. The
interaction between atoms is neglected because it is too small in
the time scale.

The atomic field operator can be decomposed to the different side
modes as
\begin{equation}
    \hat{\Psi}(\mathbf{r},t) = \sum_\mathbf{q} \psi_0(\mathbf{r}) e^{i \mathbf{q}\cdot \mathbf{r}} e^{-i\mu t}
    \hat{c}_\mathbf{q}(t).
\end{equation}
the operator $\hat{c}_\mathbf{q}$ refers to the recoiling atom wave function
which scatters the $\mathbf{q}$ mode light, $\psi_0(\mathbf{r})$ is
the ground-state wave-function of the condensate with the chemical
potential $\mu $. When the OL potential is weak enough, we could
deal this case as in a magnetic trap\cite{Moore1999prl}. And when it
is extremely strong, we are able to approximate the potential as the
spatial replication of harmonic trap $V(\mathbf{r}) = (V_0/a_0^2)
x^2$, then decompose the operator to the eigenstate of the trap that
\begin{equation}
    \hat c_\mathbf{q} = \sum_{n\neq0} \langle \psi_n|\psi_0e^{i\mathbf{q}\cdot \mathbf{r}}\rangle \hat c_n,
\end{equation}
and we define the coefficient
\begin{equation}
    A_n \equiv \langle \psi_n|\psi_0e^{i\mathbf{q}\cdot \mathbf{r}}\rangle = \sqrt{P(n,\lambda)},
\end{equation}
$\psi_n$ is the eigenfunction of $n$-th level of the trap, and the
$P(n,\lambda)$ is the Poission distribution with parameter $\lambda
= (q\sqrt{\hbar/2m\omega_T})^2 = \omega_r/\omega_T$, here $\omega_r
= \hbar q^2/2m$ is the recoiling frequency. Considering the
first-order side modes, then the interaction Hamiltonian becomes
\begin{equation}
    \hat H_i=\sum_{\mathbf{q}\neq0}\sum_{n\neq0}\int d^{3}\mathbf{k} [\hbar g(\mathbf{k})\rho_\mathbf{q}(\mathbf{k})
    A_n\hat  c_n^\dagger\hat  b^\dagger(k)\hat  c_0 + H. c.],
\end{equation}
with
\begin{equation}
\rho_\mathbf{q}(\mathbf{k})=\int d^3
\mathbf{r}|\psi_0(\mathbf{r})|^2\exp[-i(\mathbf{k}-\mathbf{k}_0+\mathbf{q})\cdot
\mathbf{r}]
\end{equation}
is the Fourier transform of the ground state density distribution
centered at $\mathbf{k}_0-\mathbf{q}$.

As the experiment shows that there are just several side modes
dominate the whole scattering process, and in order to simplify this
problem, we just take one side mode $q$ into consideration. Under
the Born-Markov approximation, the optical field could be obtained
that
\begin{equation}\label{bevo}
  \hat b(t) = \hat{b}(0)e^{-i\omega_kt}+\sum_{n\neq0}  g(\mathbf{k})\rho_\mathbf{q}(\mathbf{k})A_n \hat c_n^\dagger
  \hat c_0\delta(\omega_k).
\end{equation}
Inserting (\ref{bevo}) into the dynamic equation of atomic field,
$\dot c_n = [H,c_n]/i\hbar$, we obtain its evolution equation
\begin{equation}\label{cqevo}
    \frac{d}{dt}\hat{c}_n = A_n\frac{G_{\mathbf q}}{2}\frac{\hat c_0^\dagger\hat  c_0}{N} \sum_{m\neq0} A_m \hat c_m +
    \hat{f}^\dagger(t) -i \omega_n\hat  c_n,
\end{equation}
with the BEC's gain with the $q$-th mode as
\begin{equation}\label{Gq}
    G_{\mathbf q} = N\frac{g^2}{k_0^2}\int d^3 k |\rho_\mathbf{q}(k)|^2\delta(|k|-k_0),
\end{equation}
where we assumed that $g(\mathbf{k})$ is isotropic in the $k_x-k_z$
plane. The first term on the {\it r.h.s.} of Eq.(\ref{cqevo}) is the
gain from the condensate. The second term is the quantum
fluctuation, which has been discussed in \cite{Moore1999prl} and
does not affect the superradiant behavior in long time, and we just
take it as an initial seed. The last term on the {\it r.h.s.} of Eq.
(\ref{cqevo}) is the energy term.

In order to compare the two cases of pumping the condensate with and
without the external potential, we need to discuss equation
(\ref{cqevo}) by the mean-field approximation, replacing the field
operator $\hat c_n$ by a c-number $c_n$. By transformation
$\widetilde{c}_n = c_n\exp(-i\omega_nt)$, the equation (\ref{cqevo})
becomes
\begin{equation}\label{cpevo}
  \frac{d}{dt} \widetilde{c}_n = A_n\frac{G_{\mathbf q}}{2}\frac{\widetilde c_0^*\widetilde  c_0}{N}
  \sum_{m\neq0} A_m \widetilde  c_m e^{i(n-m)\omega_Tt}.
\end{equation}
For the case that potential is turned off when pumping the
condensate by the laser, we assume that there is just only one level
$n$ satisfying the condition $A_n=1$, with eigenenergy $\omega_n =
p^2/2m=\omega_r$, and for the else $A_n=0$. Thus on the {\it r.h.s.}
of the equation (\ref{cpevo}) the phase factor $e^{i(n-m)\omega_Tt}$
is always equal to one. And this form is consistent with the
equation (8) in~\cite{Moore1999prl}. The gain of atomic number is
therefore $G_\mathbf q$. For the case that the external potential
exists, the parameters $A_n$ are centered at $\omega_r/\omega_T$ and
have a standard deviation $\sqrt{\omega_r/\omega_T}$, and we just
need to consider these $2\sqrt{\omega_r/\omega_T}$ trap levels that
make $A_n\neq0$. And the phase factor $e^{i(n-m)\omega_Tt}$ is
different for different levels. These factors could be approximated
by calculating the average phase difference
$\omega_T\sqrt{\omega_r/\omega_T}dt = \sqrt{\omega_T\omega_r}dt$ in
a during time $dt$. Thus the different phases in different levels
lead to that the sum in the {\it r.h.s.} of equation (\ref{cpevo})
is smaller than the case that they are the same phase, in the
previous case. This results in the gain loss, given by
\begin{equation}
  G_{\mathbf q}'-G_{\mathbf q} \propto -\frac{\sqrt{\omega_T\omega_r}}{N}.
\end{equation}

A typical value of the gain is $G_\mathbf q = 4\times10^{4}$ for $I
= 100mW/cm^2$ and $N=10^6$, which is close to the frequency of the
atom in the optical lattice trap. Thus when the $\omega_T$ is large
enough to dephase the coherence of condensate and side mode, the
gain of light is suppressed. Since the effect of the trap is just a
shift in the gain, in the below we mainly discuss the gain without
the trap $G_\mathbf q$.

\section{Gain from an Array of Condensates in Released Trap}

Now we consider the case that the pump beam immediately incidents
after switching off the potential, so the external trapping
potential could be neglected. The wave function of ground state for
the $i$-th site is the Wannier function $w_i(\mathbf{r})$, which is
approximated by the gaussian function $\exp(-\sum_{j=1}^{3}
r_{j}^{2}/ 2\sigma_{j}^{2})$ with the half width of the wave
function $\sigma_{j}$ in $j$ direction $(j=x,y,z)$.  Thus the
ground-state wave-function in the optical lattice is given by $
\psi_{0}(\mathbf{r})= C_{nor}\sum w_{i}(x, y, z)$, where the
$C_{nor} = (\int d^3\mathbf{r}|\sum_{i = 1}^M
w_{i}(\mathbf{r})|^2)^{1/2}$ is the normalization factor. For one
site, we assume $\sigma_z \gg \sigma_x,\ \sigma_y$, the maximum gain
is the $z$ direction because the photon could experience the most
atomic amplification~\cite{Moore1999prl}. Here, we need to carefully
discuss the gain for the whole atomic cloud. Since the $M$ sites are
placed along $x$-axis and centered at origin with lattice spacing
$a_0$, then the atomic density can be expressed as
\begin{equation}\label{psi0}
  |\psi_0(\mathbf{r} )|^2 = C_{nor}^2[\sum_{i=1}^M|w_i(\mathbf{r})|^2
  + 2 \sum_{i=1}^{M-1}w_i(\mathbf{r}) w_{i+1}(\mathbf{r})],
\end{equation}
where the first term on the {\it r.h.s.} describes the atomic
density of site $i$. The second term is the overlapping between
neighboring sites, which is considered only when the wavefunction of
one site is wide enough to overlap its neighbors. The Fourier
transformation of the density at $\mathbf q = \mathbf k_{0}$ is
\begin{equation}
    \rho_{\mathbf k_0}(k) = C_{nor}^2 \exp(-\sum_{j=1}^{3} \frac{\sigma_j^2
k_j^2}{4})\frac{\sin(\frac{Ma_0}{2}k_x)}{\sin(\frac{a_0}{2}k_x)}[1+
     \exp(-\frac{a_0^2}{4\sigma_x^2})].
\end{equation}

\begin{figure}[tbp]
   \begin{center}
        \includegraphics[width=8cm]{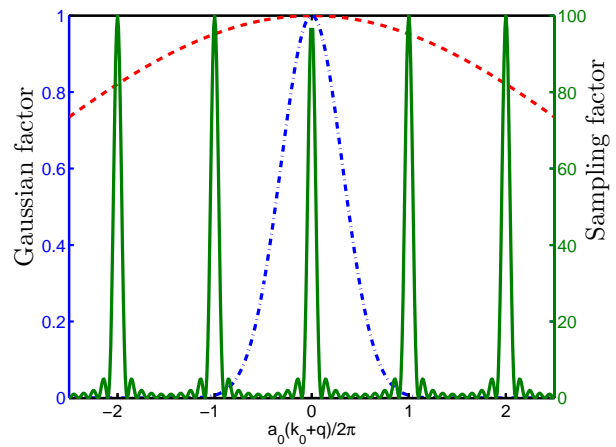}
   \end{center}
    \caption{(Color online) The sampling and the gaussian factor. The solid line is the sampling function for M=10,
     which is the result of interference between different site, and the gaussian factor is the result
     of single site amplification of light, which are drawn in dash line for $\sigma_x = 0.1a_0$, and
     in dash dotted line for $\sigma_x=a_0$.}
    \label{fig:2}
\end{figure}

We proceed to calculate the superradiant gain in (\ref{Gq}). The
factor in $\rho_\mathbf q(\mathbf k)$,
$\sin(\frac{Ma_0}{2}k_x)/\sin(\frac{a_0}{2}k_x)$, as shown in solid
line of Fig. \ref{fig:2} gives the density profile a sampling, which
means that the $\rho_\mathbf q(\mathbf k)$ is not zero in regions
with half width $2\pi/Ma_0$ and these regions separating $2\pi/a_0$
from each other. Moreover, the factor of $\rho_\mathbf q(\mathbf
k)$, $\exp(-\sum_{j=1}^{3} \frac{\sigma_j^2 k_j^2}{4})$, is centered
at $\mathbf k_0 - \mathbf q$ and spreads in $\hat{k}_i$ direction,
which is significant in the region $|k_i-k_0+q|<1/\sigma_i$. The
gaussian factor of $\rho_\mathbf q(\mathbf k)$ is the dashed and
dot-dashed line in Fig. \ref{fig:2}, which shows that it is wider
when the width in the single site wave function is smaller. We
assume that $k_0 \gg 1/\sigma_j$, $k_0 \gg 1/a_{0}$, which means
that the atomic momentum is narrow enough that all its components
could contribute to the optical amplification~\cite{Moore1999prl}.
And in the region of $\rho_\mathbf q(\mathbf k)$ where the gaussian
factor is significant, we can approximate the surface of the sphere
$ |k|=|k_0|$ as a plane tangent to the sphere, and the integral in
(\ref{Gq}) is equivalent to be the integral on this plane. So the
integral when $\mathbf{q} =\mathbf{k_0} + |k_0|\mathbf{\hat\theta}$,
where $\hat{\theta} =
\cos\theta\mathbf{\hat{k}_x}+\sin\theta\mathbf{\hat{k}_z}$, is a
unit direction in the $k_x-k_z$ plane.

Here we mainly consider the maximum gain of light in two extreme
directions, $\theta=0$ corresponding to the $z$ direction, and
$\theta = \pi/2$ for $x$ direction. For $\theta = \pi/2$, the gain
is
\begin{equation}\label{gx}
  G_{x} = G_0 \frac{M^2}{\sigma_z},
\end{equation}
where
\begin{equation}
G_0 = \frac{g^2}{k_0^2}C_{nor}^2 \frac{2\pi}{\sigma_y}[1+
  \exp(-\frac{a^2}{4\sigma_x^2})],
\end{equation}
is related to the normalization factor $C_{nor}$, the width in
$y$-direction $\sigma_y$ and the factor
$[1+\exp(-\frac{a^2}{4\sigma_x^2})]$ which reflects the coherence of
neighboring sites. Eq.(\ref{gx}) shows that the gain $G_{x}$
is proportional to the number of site squared $M^2$, which is the
result of cooperative radiation.

Because the non-zero region of $\delta$-function in (\ref{Gq}) is a
plane parallel to the $k_x-k_y$ plane when $\theta = 0$, the profile
width of $\rho_\mathbf q(k)$ will affect the result of $G_\mathbf
q$. Hence, we need to discuss two condition $\sigma_x<a_0$ and
$\sigma_x \geq a_0$. For the case that $\sigma_x<a_0$, $\rho_\mathbf
q(\mathbf k)$ has $a_0/\sigma_x$ side bands, after summing all these
side bands whose half width is $1/Ma_0$, we get the superradiant gain
given by
\begin{eqnarray}
  G_{z} = G_0 M^2\frac{a_0}{\sigma_x}\frac{1}{Ma_0}
  = G_0 \frac{M}{\sigma_x}.
\end{eqnarray}
For the case that $\sigma_x\geq a_0$, there is just one non-zero
region of $\rho_\mathbf q(\mathbf k)$, thus, the superradiant gain
is
\begin{eqnarray}
  G_{z} = G_0 M^2\frac{1}{Ma_0}
  = G_0 \frac{M}{a_0}.
\end{eqnarray}
In both cases, the gain $G_z$ is proportional to $M$, due to the
incoherent sum of different sites.

When $\sigma_x<a_0$, the gain ratio for the two extreme direction is
that $\frac{G_{x}}{G_{z}} = \frac{M a_0}{\sigma_z}=L/\sigma_z$ which
is the aspect ratio, consistent with the theory without OL trap.
When $\sigma_x>a_0$, the gain ratio becomes $\frac{G_{x}}{G_{z}} =
\frac{M\sigma_x}{\sigma_z}$ which is the effective length ratio in
this two directions. It should be noted that this theory is sound
under the condition that $k_0 \gg 1/\sigma_x$. A typical value is
$k_0 = 2\pi/780nm$, hence $\sigma_x \gg 780nm$. If we need the gain
in $z$ direction larger than that in $x$ direction, we need
$\sigma_z\gg M\cdot780nm$, which is hard to be realized in
experiment. Thus the radiation usually takes place in the $x$
direction.

\section{Spectroscopy of Superradiant Scattering}
In the previous section, we understand that the gain of light is
usually propagating along the $x$-axis. Considering $\mathbf q=
q\mathbf{\hat{x}}$ in the $x$-direction, for the different $\mathbf{q}$ the gain can be expressed as
\begin{equation}
  G_\mathbf q = \frac{G_0}{\sigma_z}\exp[-\frac{\sigma_x^2(k_0+q)^2}{2}]\frac{\sin^{2}
  \frac{Ma_0(k_0+q)}{2}}{\sin^{2}\frac{a_0(k_0+q)}{2}}.
\end{equation}

In this equation, we know that the maximum gains emerge at
$\frac{a_0(k_0+q)}{2}=n \pi$. In other words, the gain has maximum
around $k_0+2n\pi/a_0$ with separation $2\pi/a_0$.

\begin{figure}[tbp]
   \begin{center}
        \includegraphics[width=7cm]{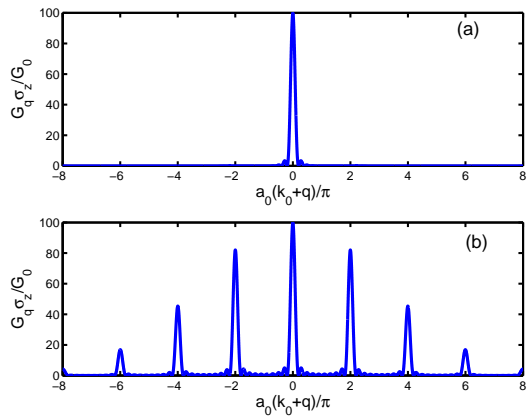}
   \end{center}
    \caption{(Color online) The gain with a factor $\sigma_z/G_0$,
     in $x$ direction for the $\sigma_x = a_0$ (a) and $\sigma_x = 0.1a_0$ (b).
    The gain has satellites when the width of single site wave function is large. When it is narrow, the
    satellites disappear. Here, $M = 10$.}
    \label{fig:3}
\end{figure}

The spectroscopy of different width of single site wave function is
plotted in Fig.~\ref{fig:3}. As shown in subfigure (a), when
$\sigma_x\geq a_0$, there is only one peak in the gain. Subfigure
(b) shows that when the $\sigma_x \ll a_0$, there are side bands.
The radiant light has side bands when the wave functions of
neighboring sites are not overlapped. The reason is that atoms in
different sites are pumping by the same phase light and become to
the same phase dipole. The radiant light which is propagating along
the lattice has a phase difference in neighboring sites, $(\mathbf
k-\mathbf k_0)\cdot a_0\mathbf{\hat x}$. Thus radiant lights with
different frequencies will have different gains by the averaging
over the whole lattice. The constructive interference will single
out the frequency component satisfying the condition $(\mathbf
k-\mathbf k_{0})\cdot a_0\mathbf{\hat x} =2n\pi$ to amplify, and
other components will be suppressed due to the destructive
interference of $M$ sites.

\begin{figure}[tbp]
   \begin{center}
        \includegraphics[width=7cm]{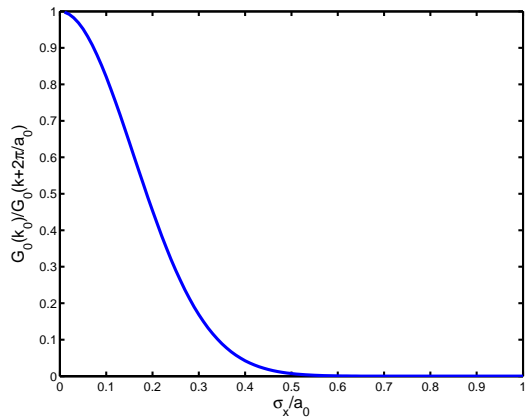}
   \end{center}
    \caption{(Color online) The ratio of maximum gain to second maximum gain in $x$ direction versus the one site
    wave packet width. Here, $M = 10$.}
    \label{fig:4}
\end{figure}

For the larger width of single site wave function, the gain of side
bands is smaller. Thus by measuring the side band gain could give us
a method to obtain the information about the width $\sigma_x$.
Fig.~\ref{fig:4} shows the ratio of maximum gain to the second
maximum gain versus $\sigma_x$. By the spectroscopical measuring, we
could obtain the information of the width of wave packet, which is
relevant to the potential quantum phase transition.

\section{Discussion and conclusions}

In the BEC superradiant experiment~\cite{Inouye1999science}, a
photon is scattered by one atom in the BEC, and this atom acquires
the recoil momentum. The moving atoms and the static BEC form a
matter wave grating, which enhances the same direction scattering.
Due to the mode competition, the highly directional emissions of
light travel along the long axis of the condensate. The stability of
the relative phase between different atomic matter waves determines
the coherence time of the matter wave. On the other hand, in the
coherent atomic recoil lasing (CARL), the situation is changed by
the presence of the cavity, the coherence is preserved as the relative
phase of the cavity light-fields, and it is independent of the
atomic motion while given by the cavity linewidth.

Different to these experimental scheme, here we extend the theory of
superradiance of BEC~\cite{Zobay2006pra} to the case of in OL trap.
In this trap, an array of atoms from the optical lattice form a
density grating, and the superradiance gain are calculated in the
quantum theory. In this theory, we consider inner-site and
inter-site coherence of atoms. Only the scattering light satisfy the
condition that $(\mathbf k-\mathbf k_{0})\cdot a_0\mathbf{\hat x}
=2n\pi$ will be singled out to be amplified and the other components
will be suppressed in different extent. Together with the grating
formed by the static and moving condensate, both gratings give
frequency selection rules. It is similar to a diode laser with
internal and external cavities. Only the light is resonant to both
cavities will be amplified.

The motion of recoiling atom in the high-frequency OL trap will
dissipate the coherence resulting in a loss of the optical gain
proportional to $\sqrt{\omega_T\omega_r}$. In the magnetic trap, the
trap frequency is smaller, and the loss can be neglected. It can
inhibit the collective radiation when the trap frequency is high
enough. By calculating the ratio of optical gain in the two extreme
direction, we show that the gain is proportional to the length that
the light travels in the condensate.

Depending on the lattice depth, the wave function for one site
overlap differently with its neighboring sites. When the OL
potential is low enough, the wave functions of neighboring sites
fully overlap, just like a condensate in the magnetic trap. When the
OL potential is high enough, the wave functions of neighboring sites
are separated. The different overlap results in the different
scattering spectroscopy. Thus the spectroscopy provides us with a
new method to detect the coherence of different sites. This
spectroscopy method offers much more precision than the absorption image
method. Moreover, unlike the time-of-flight method, which is usually
used to detect quantum phase transition~\cite{bloch}, the
spectroscopy method is a non-destructive method.  More understanding
of this mechanism can help to understand the self-organization,
especially how the long-range order arise in the
self-synchronization process. Superradiance may be helpful to detect
the phase transition between the superfluid (SF) and
Mott-insulate(MI).

\section*{ACKNOWLEDGMENTS}

We thank Dr. L. Yin to read our manuscript and give us helpful
advice. This work is partially supported by the state Key
Development Program for Basic Research of China (No.2005CB724503,
2006CB921402 and 2006CB921401),and NSFC(No.10574005 and 10874008).

\end{document}